\documentclass[prb,floatfix,twocolumn,superscriptaddress]{revtex4}

\usepackage{graphicx}
\usepackage{bm}
\usepackage{amsmath}

\newcommand{\ket}[1]{\left\vert#1\right\rangle}

\def\bK{{\bf K}} \def\bk{{\bf k}} 

  \def\bx{{\bf x}}
  \def\bAopt{{\bf A}_{\mathrm{opt}}}
\def\bE{{\bf E}} \def\bEDC{{\bf E}_{\mathrm{dc}}}

\def\bkperp{{\bf k}_\perp}

\def\Ai{\mathrm{Ai}}

\def\bzhat{\hat{\mathbf{z}}}
\def\bF{{\bf F}}
\def\bG{{\bf F}}
\def\bp{{\bf p}}
\def\EDC{E_{\mathrm{dc}}}

\def\bEopt{{\bf E}_{\mathrm{opt}}}

\def\dodh{\hat{\omega}_d}
\def\Gh{F}
\def\bV{{\bf V}}  \def\bx{{\bf x}}
\def\Heff{H_{\mathrm{eff}}}

\begin{document}
\title{Polarization dependence of the two-photon Franz-Keldysh effect}
\author{J. K. Wahlstrand}
\email{wahlstrj@umd.edu}
\affiliation{JILA, National Institute of Standards and Technology, and University of Colorado, Boulder, CO 80309 USA}
\affiliation{Department of Physics, University of Maryland, College Park, MD 20742 USA}
\author{S. T. Cundiff}
\affiliation{JILA, National Institute of Standards and Technology, and University of Colorado, Boulder, CO 80309 USA}
\author{J. E. Sipe}
\altaffiliation[Permanent address: ]{Department of Physics, University of Toronto, Toronto, Ontario, Canada M5S1A7}
\affiliation{JILA, National Institute of Standards and Technology, and University of Colorado, Boulder, CO 80309 USA}
\date{\today}

\begin{abstract}
The effect of a constant electric field on two-photon absorption in a direct band gap semiconductor is calculated using an independent-particle theory.
Two band structure models for GaAs are used:  a two-band parabolic model and an eight-band $\bk \cdot\bp$ model.
Both predict a strong dependence of the two-photon electroabsorption spectrum on the polarization of the light with respect to the constant field.
We attribute the polarization dependence to the strong effect of a constant field on intraband dynamics.
\end{abstract}

\maketitle

Multiphoton absorption is an important topic in the study of light-matter interaction, and will only become more important with our increasing ability to control the spectral and quantum statistical properties of light.
Pulse shaping,\cite{meshulach_coherent_1998} use of entangled photons,\cite{dayan_nonlinear_2005} and quantum interference between absorption pathways\cite{shapiro_coherent_2003} have all been found to affect multiphoton absorption processes in profound and useful ways.
Nonlinear absorption has also become an increasingly important issue in photonics.\cite{lin_nonlinear_2007,ahmad_energy_2008}
Despite strong interest in this area, the effects of constant electric fields have not received much attention, which is surprising considering that strong fields are often present in semiconductor-based photonic devices.
Field-induced changes in the optical absorption in a bulk semiconductor are predominantly caused by the Franz-Keldysh effect (FKE),\cite{keldysh__1958,franz__1958,aspnes_electric_1967} related to the acceleration of carriers by a constant (dc) electric field.
In the case of linear absorption, the field induces an exponential tail for photon energies below the band gap caused by field-induced tunneling, and oscillations for energies above the band gap, related to the coherence of carriers.

The two-photon FKE was the subject of a number of theoretical studies decades ago.\cite{yang_two-photon_1971,yang_electric-field_1973-1,yang_electric-field_1973,hassan_two-photon_1975,kolber_theory_1978,areshev_2-photon_1979}
There has recently been a resurgence of interest, with theoretical works on bulk semiconductors by Garcia and Kalyanaraman\cite{garcia_tunneling_2006,garcia_phonon-assisted_2006} and in nanostructures by Xia and Spector.\cite{xia_franzkeldysh_2009,xia_nonlinear_2010}
Both used two-band models in the parabolic band approximation (PBA) to find analytical expressions for the two-photon absorption coefficient in the presence of an electric field.
Experimental results for photon energies below the band gap have also recently been published.\cite{cui_modulation_2008}
Here we extend a recently described theory\cite{wahlstrand_theory_2010} of the one-photon FKE in bulk semiconductors to two-photon absorption.
We calculate two-photon electroabsorption (EA), the difference between the absorption spectrum in the presence and in the absence of a dc field, using an eight-band $\bk \cdot \bp$ model for GaAs and find that the results agree qualitatively with analytical expressions from a two-band PBA model.
The theory predicts a strong dependence of the two-photon EA spectrum on the polarization of the optical field with respect to the dc field.
We give expressions for two-photon absorption with optical pulses, enabling calculation of effects that depend on the pulse shape,\cite{meshulach_coherent_1998} and simpler expressions suitable for long pulses.

We consider a semiconductor in the presence of a uniform, constant electric field $\bEDC$.
Much of the theory is the same as for one-photon absorption;\cite{wahlstrand_theory_2010} here we only sketch the approach and give the relevant equations.
In the independent-particle approximation, neglecting the interaction between carriers and scattering processes, the Hamiltonian including the effect of the dc field is
\begin{equation}
\mathcal{H}_{\mathrm{dc}}(t)=\frac{1}{2m}\left( \frac{\hbar }{i}{\bm\nabla }+ \hbar \bK(t)  \right) ^{2}+V(\mathbf{x}), \nonumber
\end{equation}
where $\bK (t) = e \bEDC t/\hbar$ describes the acceleration of electrons by the electric field and $V(\bx)$ describes the interaction of electrons with the ions of the crystal.
We use instantaneous eigenstates $\{\bar{\phi}_n (\bk; \bx)\}$ of $\mathcal{H}_{\mathrm{dc}}(t)$ as basis states,\cite{wahlstrand_theory_2010,sipe_nonlinear_1993} denoting velocity matrix elements $\bV_{mn}(\bk; t)$, electric dipole matrix elements $\bm{\mu}_{mn} (\bk; t)$, and band energies $\omega_n(\bk)$.
We denote destruction operators for these states as $b_{n\bk}$.
To account for coupling between bands due to the dc field, we calculate absorption using states $\{\bar{\chi}_n (\bk; \bx)\}$ that are related to $\{\bar{\phi}_n (\bk; \bx)\}$ by a unitary evolution matrix $\mathrm{L}(\bk; t)$ satisfying
\begin{equation}
i\hbar \frac{d\mathrm{L}\left( \mathbf{k};t\right) }{dt}=\left[ \mathrm{T}(%
\mathbf{k};t)+\mathrm{S}(\mathbf{k};t)\right] \mathrm{L}\left( \mathbf{%
k};t\right) ,  \nonumber
\end{equation}
where $T_{pm}(\mathbf{k};t) =\delta _{pm}\hbar \omega _{m}\bm{(}\mathbf{k}+\mathbf{K}%
(t)\bm{)}$ and $S_{pm}(\mathbf{k};t) =-\bm{\mu }_{pm}(\mathbf{k};t)\cdot \mathbf{E}_{%
\mathrm{dc}}$.

It is assumed that the nominal dc field is turned on before the optical pulse arrives.
Optical absorption is calculated in an interaction picture, where the ket $\left\vert \Psi (t)\right\rangle $ evolves according to the interaction Hamiltonian $%
H_{\mathrm{eff}}(t)$, 
\begin{equation}
i\hbar \frac{d\left\vert \Psi (t)\right\rangle }{dt}=H_{\mathrm{eff}}(t)\left\vert \Psi (t)\right\rangle .
\label{SE}
\end{equation}
Here $H_{\mathrm{eff}}(t)$ describes the effect of the optical field $\bEopt(t) = -(1/c) d\bAopt/dt$ in the presence of the dc field; it is given by\cite{wahlstrand_theory_2010}
\begin{equation}
H_{\mathrm{eff}}(t)=-\frac{e}{c}\sum_{n_{1},n_{2},\mathbf{k}}b_{n_{2}\mathbf{k}}^{\dagger }b_{n_{1}\mathbf{k}} \left[ \mathbf{A}_{\mathrm{opt}}(t)\cdot \mathbf{\tilde{V}}_{n_{2}n_{1}}(\mathbf{k};t) \right], \nonumber
\end{equation}
where 
\begin{equation}
\mathbf{\tilde{V}}_{nq}(\mathbf{k};t)=\sum_{m,p}L_{mn}^{\ast }(\mathbf{k};t) \mathbf{V}_{mp}(\mathbf{k};t)L_{pq}(\mathbf{k};t).
\label{Vtilde}
\end{equation}
The initial condition of Eq.~(\ref{SE}) $\left\vert \Psi ^{H}\right\rangle$ is the state of the system at the moment the dc field is turned on.\cite{wahlstrand_theory_2010}
The matrix elements $\mathbf{\tilde{V}}_{nq}(\mathbf{k};t)$ appearing in $\Heff(t)$ are effective matrix elements that take into account the fact that the dc field can induce transitions between these bands itself.
We assume an optical pulse of the form
\begin{equation}
\bEopt(t)=\int \frac{d\omega }{2\pi }\mathbf{E}(\omega)e^{-i\omega t},
\label{EFT}
\end{equation}
and assume that $\bE(\omega)$ does not contain frequency components above the band gap so one-photon absorption may be neglected.

We calculate $\ket{\Psi^{(2)}}\equiv\ket{\Psi^{(2)}(\infty)}$ in a perturbative solution of the Schr\"{o}dinger equation [Eq.~(\ref{SE})], \emph{i.e.} 
\begin{math}
\left\vert \Psi (t)\right\rangle =\left\vert \Psi ^{H}\right\rangle
+\left\vert \Psi ^{(1)}(t)\right\rangle + \ket{\Psi^{(2)} (t)} + \cdots ,  \nonumber
\end{math}
which is
\begin{equation}
\left\vert \Psi ^{(2)}\right\rangle =\frac{1}{(i\hbar)^2 }\int_{-\infty}^{\infty} \Heff (t^\prime) \int_{-\infty
}^{t^\prime}H_{\mathrm{eff}}(t^{\prime \prime})\left\vert \Psi ^{H}\right\rangle
dt^{\prime\prime} dt^{\prime}.
\nonumber
\end{equation}
The number of carriers injected in some volume is $\Delta N = \left\langle \Psi^{(2)} \vert \Psi^{(2)} \right \rangle$.
For the carrier \emph{density} injected $\Delta n$, we find
\begin{multline}
\Delta n  \label{2pDC} =\int_{0}^{\infty }d\omega _{a}\int d\omega _{d}d\omega _{d}^{\prime
} \eta^{ijlm} \left(\omega_a,\omega_d,\omega_d^\prime\right)  \\ \times E^{i}\left(\omega _{a}-\frac{1}{2}\omega _{d}\right)E^{j}\left( \omega _{a}+%
\frac{1}{2}\omega _{d}\right) \\ \times \left[ E^{l}\left(\omega _{a}-\frac{1}{2}\omega
_{d}^{\prime }\right)E^{m}\left( \omega _{a}+\frac{1}{2}\omega _{d}^{\prime
}\right) \right] ^{\ast },
\end{multline}
where $i,j,l,m$ denote vector components and we have defined a carrier injection tensor
\begin{multline}
\eta^{ijlm} (\omega_a,\omega_d,\omega_d^\prime) \\= \frac{\varepsilon }{2} \sum_{cv}\int \frac{d\mathbf{k}_{\perp }}{4\pi
^{2}} \theta _{cv\mathbf{k}_{\perp }}^{ij}(\omega _{a},\omega
_{d}) \left[ \theta _{cv\mathbf{k}_{\perp }}^{lm}(\omega _{a},\omega
_{d}^{\prime })\right] ^{\ast },
\label{eta}
\end{multline}
in which
\begin{multline}
\theta^{ij}_{cv\bk_\perp} (\omega_a,\omega_d) = \frac{ie^2}{\pi\hbar^2(4\omega_a^2-\omega_d^2)} \\ \times \sum_n \int d\dodh \left[ \frac{\Gh^i_{cn}(\bkperp; -\omega_a + \frac{1}{2}\dodh) \Gh^j_{nv} (\bkperp; -\omega_a - \frac{1}{2}\dodh)}{\omega_d - \dodh}\right. \\ \left.+\frac{\Gh^j_{cn}(\bkperp; -\omega_a - \frac{1}{2}\dodh) \Gh^i_{nv} (\bkperp; -\omega_a + \frac{1}{2}\dodh)}{\omega_d - \dodh} \right],
\label{theta2}
\end{multline}
where, finally,
\begin{equation}
\bF_{mn} (\bkperp; -\omega) = \int \tilde{\bV}_{mn} (\bkperp; t) e^{i\omega t}dt.
\label{Fomega}
\end{equation}

Equation (\ref{2pDC}) is relevant for a pulsed optical field, including effects due to pulse shaping.\cite{meshulach_coherent_1998}
In the limit of a long optical pulse with a constant optical field envelope $\bE$, one can derive a Fermi golden rule (FGR) expression for the rate of carrier injection
$\dot{n} = 16\pi^3 \eta^{ijlm} (\omega, 0, 0) E^i E^j [E^l E^m]^*$.
The third-order nonlinear susceptibility is given by $\mathrm{Im}[\chi^{(3)}(-\omega; -\omega,\omega,\omega)]=(16\pi^3\hbar/3)\eta (\omega, 0, 0)$.
For the dc field pointing along a crystal direction, the nonlinear absorption coefficient $\beta(\omega)$ is\cite{sutherland_handbook_1996}
\begin{equation}
\beta (\omega) = \frac{128\pi^5\hbar\omega}{n^2(\omega)c^2} \eta^{iiii} (\omega,0,0),
\end{equation}
(no summation implied) for light linearly polarized in the $i$ direction.

For two parabolic bands and a $\bk$-independent interband dipole matrix element $\bV_{cv}$, the equations can be solved analytically, and the two-photon absorption spectrum may be expressed in terms of Airy functions.
We assume that the bands have the form
\begin{math}
\hbar \omega_{c} (\bk)= \hbar \omega_g + \hbar^2 k^2/(2m_c)
\end{math}
and
\begin{math}
\hbar \omega_{v} (\bk) = -\hbar^2 k^2/(2m_h)
\end{math},
where $m_c$ is the electron mass, $m_h$ is the hole mass, and $\hbar\omega_g$ is the band gap.
The energy difference between the valence and conduction bands is
\begin{math}
\hbar \omega_{cv} (\bk) = \hbar\omega_g+ \hbar^2 k^2/(2\mu),
\end{math}
where $\mu=(m_c^{-1} + m_h^{-1})^{-1}$ is the reduced mass.
We assume a dc field pointing in the $\bzhat$ direction and explicitly denote perpendicular and parallel components of the wavevector, $\bk=\bkperp+k_\parallel \bzhat$.
The off-diagonal components are 
\begin{equation}
\bG_{cv}(\bk_\perp; -\omega) = \bV_{cv} \frac{1}{\Omega} \Ai\left(-\frac{\omega-\omega_g-\hbar k_\perp^2/2\mu}{\Omega}\right),
\label{money}
\end{equation}
where $\Ai(x)$ is the Airy function and the electro-optic frequency $\Omega = (\hbar \varepsilon^2/2\mu)^{1/3}$, defining a normalized dc field $\varepsilon\equiv (e/\hbar) \EDC$.
Using $\bV_{nn} (\bk) = \hbar \bk/m_n$ and $\bV_{nn} (\bk; t) = \bV_{nn} (\bkperp; t+k_\parallel/\varepsilon)$,\cite{wahlstrand_theory_2010} we have $\bV_{nn} (\bk_\perp;t) = \hbar \bk_\perp/m_n + \hat{\mathbf{z}}\hbar \varepsilon t/m_n$.
Using this in Eq.~(\ref{Vtilde}) and Eq.~(\ref{Fomega}) and using the fact that $\mathrm{S}=0$ in the absence of interband coupling,\cite{wahlstrand_theory_2010} we find
\begin{eqnarray}
\bG_{nn}(\bk_\perp; -\omega) &=& \frac{\hbar \bk_\perp}{m_n} \delta(\omega) + \hat{\mathbf{z}} \frac{ i\hbar \varepsilon}{m_n} \delta'(\omega),
\label{gammadiag}
\end{eqnarray}
where we have used $t e^{-i\omega t} = id(e^{-i\omega t})/d\omega$.
We see that the diagonal (intraband) matrix element is qualitatively different in the direction of the dc electric field $\bzhat$, not surprising considering that the electric field accelerates carriers.

We find, using Eqs.~(\ref{money}) and (\ref{gammadiag}) in Eq.~(\ref{theta2}),
\begin{multline}
\theta^{zz}_{cv\bk_\perp} (\omega_a,\omega_d) = \frac{8 e^2\varepsilon}{\pi \hbar\Omega^2\mu(4\omega_a^2-\omega_d^2)^3}V^z_{cv} \\ \times \left[ \Omega(4\omega_a^2+\omega_d^2) \Ai' \left(-\frac{2\omega_a-\omega_g-\hbar k_\perp^2/2\mu}{\Omega} \right) \right. -\\ \left.  \omega_a (4\omega_a^2-\omega_d^2) \Ai\left(-\frac{2\omega_a-\omega_g-\hbar k_\perp^2/2\mu}{\Omega}\right)\right],
\label{thetazz}
\end{multline}
and
\begin{multline}
\theta^{xx}_{cv\bk_\perp} (\omega_a,\omega_d) = \\ \frac{-8ie^2 k^x \omega_a}{\pi \hbar\Omega\mu (4\omega_a^2-\omega_d^2)^2}V^x_{cv} 
  \Ai\left(-\frac{2\omega_a-\omega_g-\hbar k_\perp^2/2\mu}{\Omega}\right). 
\label{thetaxx}
\end{multline}
The $y$-component is simply the above with $x\rightarrow y$.

\begin{figure}
\includegraphics[width=8.0cm]{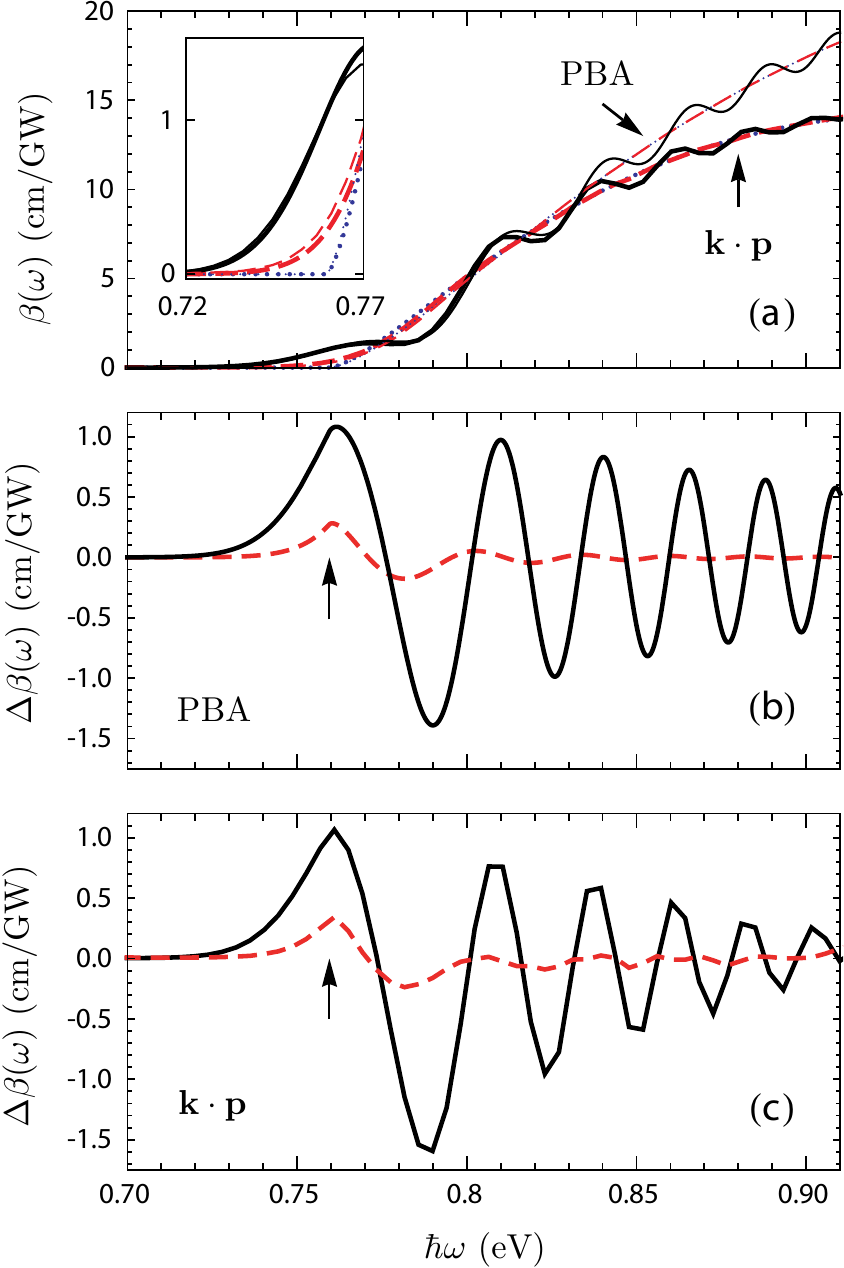}
\caption{(color online)  Calculated two-photon Franz-Keldysh effect in GaAs for $\EDC = 67$ kV/cm.  Blue dotted:  no dc field; red dashed:  $\bEopt \perp \bEDC$; black solid: $\bEopt \parallel \bEDC$.  (a) Two-photon absorption spectrum $\beta(\omega)$ using a model with two bands in the parabolic band approximation (PBA, thin lines) and an eight-band $\bk \cdot\bp$ model (thick lines).  The inset shows a close-up view of the absorption spectrum near the half bandgap $\hbar \omega_g/2$, with the units on the two axes the same as the main plot.  (b,c) Two-photon electroabsorption spectrum [$\beta(\omega)$ with a dc field minus $\beta(\omega)$ without a field]. The half band gap is marked with an arrow.  (b) Two-band PBA model.  (c) Eight-band $\bk \cdot \bp$ model.}
\label{results}
\end{figure}

Using Eqs.~(\ref{thetazz}) and (\ref{thetaxx}) in Eq.~(\ref{eta}) we find in the FGR limit, for the optical field parallel to the dc field,
\begin{multline}
\eta^{zzzz} (\omega, 0, 0) = \frac{e^4 \varepsilon^3}{16\pi^3\hbar^3 \omega^6\Omega\mu} |V^z_{cv}|^2 \\ \times \left\{ \left[ \frac{(\omega_g-2\omega)^2}{3\Omega^4}-\frac{1}{2\Omega\omega} -\frac{\omega_g-2\omega}{\Omega\omega^2}\right]\Ai^2\left(\frac{\omega_g-2\omega}{\Omega}\right)  \right. \\ - \left. \frac{2}{3\Omega^2} \Ai'\left(\frac{\omega_g-2\omega}{\Omega}\right) \Ai\left(\frac{\omega_g-2\omega}{\Omega}\right) \right. \\ \left. + \left[ \frac{1}{\omega^2}-\frac{\omega_g-2\omega}{3\Omega^3}\right] \left[\Ai'\left(\frac{\omega_g-2\omega}{\Omega}\right)\right]^2 \right\} ,
\label{pba_perp}
\end{multline}
whereas for the optical field perpendicular to the dc field we have
\begin{multline}
\eta^{xxxx}(\omega, 0, 0) = \frac{e^4\varepsilon}{48\pi^3\hbar^4\omega^6} |V^x_{cv}|^2 \\ \times \left\{ -2\frac{\omega_g-2\omega}{\Omega}\left[\Ai'\left(\frac{\omega_g-2\omega}{\Omega}\right)\right]^2 \right. \\ \left. +2\left(\frac{\omega_g-2\omega}{\Omega}\right)^2\Ai^2\left( \frac{\omega_g-2\omega}{\Omega}\right)\right.\\ \left.-\Ai\left(\frac{\omega_g-2\omega}{\Omega}\right)\Ai'\left(\frac{\omega_g-2\omega}{\Omega}\right)\right\}.
\label{pba_parallel}
\end{multline}
In the limit of no dc field, one can derive using asymptotic expressions for $\Ai(x)$ that Eqs.~(\ref{pba_perp}) and (\ref{pba_parallel}) both reduce to the well-known expression\cite{pidgeon_two-photon_1979} for allowed-forbidden transitions between parabolic bands
\begin{equation}
\eta^{iiii}(\omega, 0, 0) = \frac{\sqrt{2}e^4\mu^{1/2}}{24\pi^4\hbar^{9/2}\omega^6} |V^i_{cv}|^2 (2\omega-\omega_g)^{3/2}.
\end{equation}
The two-photon absorption coefficient for the two polarization configurations is plotted in Fig.~\ref{results}.
Results from the PBA model are shown in Fig.~\ref{results}a, assuming $\hbar V^i_{cv} = 10.3$ eV$\cdot \mathrm{\AA}$ and the light hole reduced mass in GaAs.
We have multiplied the PBA results by 2 to account for the spin degeneracy not included in the model.

Note that the expression for fields perpendicular ($\eta^{xxxx}$) is the same as the one derived by Garcia,\cite{garcia_tunneling_2006} but the one for fields parallel ($\eta^{zzzz}$) is very different.
This difference is especially obvious in the two-photon electroabsorption spectrum, shown in Fig.~\ref{results}b.
At a fundamental level this is not surprising: \ Although the FKE goes beyond perturbation theory, consider for the moment a simple picture where the $\eta ^{ijkl}$ arise from a perturbation due to the dc field of the imaginary part of $\chi _{(3)}^{ijkl}(-\omega ;-\omega ,\omega ,\omega )$ that describes two-photon absorption.
For light polarized along the $x$ and $z$ directions the appropriate components are $\chi_{(3)}^{xxxx}$ and $\chi_{(3)}^{zzzz}$ respectively, which are obviously equal for a cubic crystal such as GaAs.
Since the band structure models we use here neglect the lack of inversion symmetry in the crystal, the leading order correction of these $\chi _{(3)}^{ijkl}$ due to a dc field involve two powers of the field.
If we let $\Gamma ^{ijklmn}$ refer formally to taking the second derivative of $\chi _{(3)}^{ijkl}(-\omega ;-\omega ,\omega ,\omega )$ with respect to an applied dc field, with components in the $m$ and $n$ direction, then since we take our electric field in the $\mathbf{\hat{z}}$ direction, in a perturbative description we could identify $\Gamma ^{xxxxzz}\leftrightarrow \eta ^{xxxx}$ and $\Gamma^{zzzzzz}\longleftrightarrow \eta ^{zzzz}$.
Even within this simple picture, because of the high rank of $\Gamma $ we would have no reason to expect $\Gamma ^{xxxxzz}$ and $\Gamma ^{zzzzzz}$ (and thus $\eta ^{xxxx}$ and $\eta ^{zzzz}$) to be
equal, especially since the frequencies associated with the last two components of the $\Gamma$'s are drastically different than those of the others.

Two-photon absorption in GaAs for energies near the half band gap is mostly due to two-band processes, and the two-band model can be expected to capture the physics of that case.\cite{catalano_interband_1988}
However, three-band processes that involve an intermediate band also contribute to two-photon absorption.\cite{pidgeon_two-photon_1979}
To examine how the polarization dependence changes when intermediate bands are included, we calculated the two-photon EA spectrum using a $\bk \cdot \bp$ model\cite{lax_symmetry_1974} for the band structure.
The model directly includes eight bands:  the six uppermost valence bands and the two lowermost conduction bands.
The spin splitting due to the lack of a center of inversion symmetry in GaAs is not accounted for in the eight-band model, so there is spin degeneracy at all $\bk$.
The coupling $P_0$ between the valence bands and conduction bands produces some of the band curvature.
Effects of remote bands on the valence band curvature are handled through modified Luttinger parameters. 
A remote band parameter $F$ fixes the conduction band effective mass to the observed value.
The model parameters are the same as was used recently in a calculation of the one-photon FKE.\cite{wahlstrand_theory_2010}
The matrix elements and the evolution matrix are calculated as in the one-photon calculation.\cite{wahlstrand_theory_2010}
We show the results of the $\bk \cdot\bp$ calculation in Fig.~\ref{results}a and \ref{results}c.
Overall, the results are in qualitative agreement with the two-band PBA model.
The difference in magnitude can be traced to the more accurate treatment of the effective mass of holes in the $\bk \cdot \bp$ model.

In summary, we have calculated the two-photon FKE and find a strong polarization dependence that we attribute to the strong effect of a dc field on intraband dynamics.
A more sophisticated model of the band structure, such as a 14-band model\cite{pfeffer_five-level_1996} recently used for the one-photon FKE,\cite{wahlstrand_theory_2010} should in addition show effects that are odd in the dc field because of the lack of inversion symmetry in GaAs.
Other future extensions to the theory could include adding the Coulomb interaction\cite{yang_electric-field_1973-1,kolber_theory_1978,areshev_2-photon_1979} and scattering.
Higher-order multi-photon absorption could be calculated by going to higher order in $\Heff$.\cite{quic_footnote}

Previous experimental work\cite{cui_modulation_2008} on the two-photon FKE only employed a dc field perpendicular to the optical field, and only measured the effect for $\hbar\omega$ below the half band gap.
A transverse geometry is required to observe the polarization dependence we predict, and that will require application of an external field,\cite{wahlstrand_uniform-field_2010} rather than the usual use of the built-in field in a doped heterostructure.
The prediction of a strong polarization dependence of the two-photon FKE, with very strong Franz-Keldysh oscillations for all fields parallel, should stimulate more interest both in experiments and in more sophisticated theories of nonlinear absorption in external fields.

J.K.W.~acknowledges support from the Joint Quantum Institute.
S.T.C.~is a staff member in the NIST Quantum Physics division.
J.E.S.~acknowledges support from the Natural Sciences and Engineering Research Council of Canada.


\end{document}